\documentclass[sigconf,authorversion,nonacm]{acmart}

\AtBeginDocument{%
  \providecommand\BibTeX{{%
    \normalfont B\kern-0.5em{\scshape i\kern-0.25em b}\kern-0.8em\TeX}}}

\setcopyright{none}

\usepackage{listings}

\usepackage{tikz}
\usetikzlibrary{calc,matrix}
\usepackage{amsthm}
\usepackage{colortbl}
\usepackage{pgfplots}
\usepackage{balance}

\definecolor{mygreen}{rgb}{0,0.6,0}
\definecolor{mygray}{rgb}{0.5,0.5,0.5}
\definecolor{mymauve}{rgb}{0.58,0,0.82}

\begin{document}

\title{Masked Matrix Multiplication for Emergent Sparsity}

\author{Brian Wheatman}
\email{wheatman@cs.jhu.edu}
\affiliation{%
  \institution{Johns Hopkins University}
  \country{}
  \city{}
}

\author{Meghana Madhyastha}
\email{mmadhya1@jhu.edu}
\affiliation{%
  \institution{Johns Hopkins University}
  \country{}
  \city{}
}

\author{Randal Burns}
\email{randal@cs.jhu.edu}
\affiliation{%
  \institution{Johns Hopkins University}
  \country{}
  \city{}
}


\begin{abstract}


  Artificial intelligence workloads, especially transformer models, exhibit emergent sparsity in which computations perform selective sparse access to dense data. The workloads are inefficient on hardware designed for dense computations and do not map well onto sparse data representations.
  We build a vectorized and parallel matrix-multiplication system $(A \times B = C)$ that eliminates unnecessary computations and avoids branches based on a runtime evaluation of sparsity. We use a combination of dynamic code lookup to adapt to the specific sparsity encoded in the B matrix and preprocessing of sparsity maps of the A and B matrices to compute conditional branches once for the whole computation.
  For a wide range of sparsity, from 60\% to 95\% zeros, our implementation performs fewer instructions and increases performance when compared with Intel MKL’s dense or sparse matrix multiply routines. Benefits can be as large as 2 times speedup and 4 times fewer instructions.
  %


\end{abstract}





\maketitle

%

\section{Introduction}




Recent results have observed that sparsity is an emergent phenomena in AI and leveraging this sparsity is a powerful tool in reducing computational cost. The Lazy Neuron Phenomena~\cite{Li2023lazy} shows that the activations of ReLU in multi-layer perceptrons in transformer models have greater than 90\% sparsity. They also observed that all neurons are active for some activations. Essentially, the data are dense, all neurons are needed, and their usage is sparse. Contextual sparsity~\cite{liu2023deja} shows that using activations to dynamically select a sparse subset of a model can approximate the output of a dense model when using only 5\% of MLP parameters and 20\% of attention heads. The potential savings are huge and neither approach realizes speedup proportional to sparsity. The problem lies in mapping sparse computation to the dense matrix multiplication primitives supported by hardware.
The Lazy Neuron authors ``emphasize that this result is far from fully realizing the benefit of sparse activation, due to a lack of hardware support for sparse computation.''~\cite{Li2023lazy}


We present a matrix multiplication system that better maps sparse computations to hardware designed for dense, vectorized computation. We call this Masked Matrix Multiplication (MMM), because it operates on data that are stored densely but are selected sparsely at runtime. Specifically, $M(X) \times N(Y)=C$ where $X$ and $Y$ are dense matrices and $M$ and $N$ are masking functions that zero out large fractions of the input and are evaluated at runtime.  For ease of presentation, we define $A=M(X)$ and $B=N(Y)$.   To be effective, MMM requires sparsity in both matrices and block-sparsity in $B$. Block sparsity means that there are small, dense sequential blocks of non-zeros on which we perform vector operations. If only one matrix is sparse, simpler techniques, such as Intel MKL's sparse-dense approach are more efficient.

The system works best for the types of sparsity observed in transformer models and at the fraction of non-zeros 60-95\% seen in the data.
The specific combination of bit-level and block-sparsity occurs in the Deja Vu system~\cite{liu2023deja}.
For the most part,
these dynamic sparse-sparse multiplications have not been explored, because they are no more efficient than sparse-dense or dense-dense multiplications without our techniques. We show that these computations exist and that they can be leveraged to improve performance. We expect that as machine-learning workloads continue to explore sparsity, MMM will become a  tool to reduce cost, power, and time.
Sparsity exists or can be leveraged in many settings, including switch transformers~\cite{Fedus2021switch}, sparse transformers~\cite{jaszczur2021sparse}, and reformers with sparse attention~\cite{Kitaev2020reformer}.

We use dynamic code execution to adapt to the sparsity encoded in the $B$ matrix and single-pass preprocessing of $A$ and $B$ to minimize branching instructions. The challenge lies in that there is sparsity in both matrices, also called dual-sided sparsity~\cite{wang2021dual}. With a single sparse matrix, one can just skip elements of that sparse matrix in the outer loop and perform dense computations in the inner loops. Our system deals with sparsity at a finer-grain in inner loops by selecting code from a lookup table that matches the block-sparse patterns seen during preprocessing. Preprocessing is parallel, cache efficient, and vectorizable.

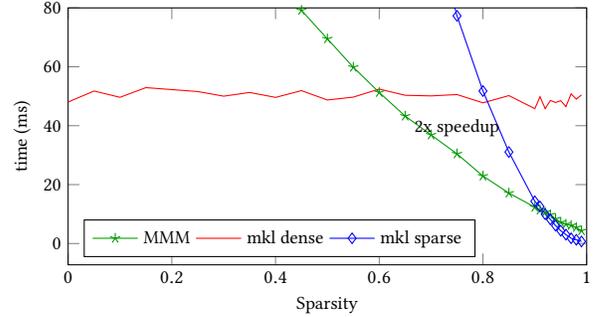
\begin{figure}
  \centering
  \footnotesize
  \begin{tikzpicture}
    \begin{axis}[ymax = 80, mark=none, height=5cm,
        width=\columnwidth,        legend pos=south west,
        legend columns=3,        xlabel = Sparsity,
        ylabel = time (ms),ylabel near ticks, xmin = 0, xmax = 1, xlabel near ticks]
      \addplot[mark size=2pt, mark=star, mygreen] table [x=algorithm, y expr=\thisrow{MMM}/1000, col sep=tab] {figures/2048_parallel_laptop.tsv};
      \addlegendentry{MMM}
      \addplot[mark size=2pt, mark=circle, red] table [x=algorithm, y expr=\thisrow{mkl dense}/1000, col sep=tab] {figures/2048_parallel_laptop.tsv};
      \addlegendentry{mkl dense}
      \addplot[mark size=2pt, mark=diamond, blue] table [x=algorithm, y expr=\thisrow{mkl sparse}/1000, col sep=tab] {figures/2048_parallel_laptop.tsv};
      \addlegendentry{mkl sparse}

      \node [above] at (axis cs:  .75,  34) {2x speedup};
    \end{axis}
  \end{tikzpicture}

  \vspace{-5pt}

  \caption{Runtime for a 2048x2048 matrix multiplication with varying sparsity.  MMM outperforms Intel MKLs best algorithm for intermediate levels of sparsity, providing 2 times speedup at 80\% zeros.}
  \label{fig:time-overview}

  \vspace{5pt}

\end{figure}

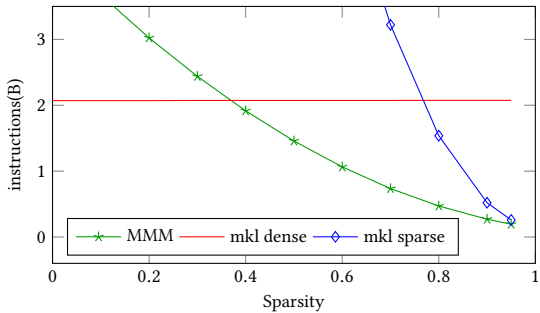
\begin{figure}
  \centering
  \footnotesize
  \begin{tikzpicture}
    \begin{axis}[ymax=3.5, ymin=-.4, mark=none, height=5cm,
        width=8cm,        legend pos=south west,
        legend columns=3,        xlabel = Sparsity,
        ylabel = instructions(B),ylabel near ticks, xmin = 0, xmax = 1, xlabel near ticks]
      \addplot[mark size=2pt, mark=star, mygreen] table [x=sparsity, y expr=\thisrow{MMM}/1000000000, col sep=tab] {figures/2048_laptop_instructions.tsv};
      \addlegendentry{MMM}
      \addplot[mark size=2pt, mark=circle, red] table [x=sparsity, y expr=\thisrow{mkl dense}/1000000000, col sep=tab] {figures/2048_laptop_instructions.tsv};
      \addlegendentry{mkl dense}
      \addplot[mark size=2pt, mark=diamond, blue] table [x=sparsity, y expr=\thisrow{mkl sparse}/1000000000, col sep=tab] {figures/2048_laptop_instructions.tsv};
      \addlegendentry{mkl sparse}

    \end{axis}
  \end{tikzpicture}

  \vspace{-5pt}

  \caption{Number of instructions to perform a 2048x2048 matrix multiplication with varying sparsity. MMM reduces instructions be a factor of 4 at 70\% zeros.}
  \label{fig:inst-overview}
\end{figure}

Overall, the system provides speedup over Intel MKL's sparse-sparse, sparse-dense, and dense-dense matrix multiplication for a wide-range of sparsity (Figure \ref{fig:time-overview}), exhibiting two times speedup at 80\% sparsity in both matrices. This is a summary result in that there are many parameters that affect performance, including $A$ sparsity, $B$ sparsity, matrix size, and architecture. The total number of instructions shows how MMM avoids computation on 0s. MMM can perform up to 4 times fewer instructions than MKL sparse or dense (Figure \ref{fig:inst-overview}). The combination of results also indicates that Intel MKL is realizing higher instructions per cycle (IPC) than MMM.
This paper will not answer the question: Can the gap between performance and instructions be narrowed in MMM with respect to Intel MKL?
We believe that the sophisticated locality, blocking, and register management used in Intel MKL \cite{mkl} are compatible with MMM and would close this gap.
However, our implementation does not explore this and a much greater engineering effort would be needed.

We present results for Intel CPUs only. GPUs are largely preferred for AI training and a GPU implementation remains as future work.
GPUs have the same vector computing and sequential memory properties that MMM leverages.
CPUs remain important for inference: 70 percent of inference in the datacenter, including the hyperscalers and cloud builders, runs on Intel Xeon CPUs~\cite{nextplatformcpu, intelcpuoptmanual}. Inference at the IoT edge largely uses CPUs~\cite{skyinference}.


\section{Background}
We describe the matrix multiplication on which MMM builds, starting
with the simple implementation of dense multiplication for row-major matrices. The basic algorithm (Figure \ref{fig:mat-simple}) uses three-nested loops that iterate over the inputs $A$ and $B$ sequentially and writes partial sums to the
output matrix $C$ sequentially.

\begin{figure}
  \centering
  \begin{lstlisting}[language=c++, basicstyle=\footnotesize]
void MatrixMultiplySimple(Matrix &C, 
    const Matrix &A, const Matrix &B) {
  for (int i = 0; i < C.rows; ++i) 
    for (int k = 0; k < B.rows; ++k) 
      for (int j = 0; j < C.cols; ++j) 
        C[i, j] += A[i, k] * B[k, j];
}

\end{lstlisting}
  \caption{Simple dense matrix multiplication.}
  \label{fig:mat-simple}
\end{figure}

Recursive calls to the simple algorithm convert the computation into a cache-oblivious algorithm that fits into all levels of cache simultaneously. This breaks each matrix into 4 sub-matrices and performs 8 matrix multiplications among the sub-matrices as seen in Figure \ref{fig:mat-recursive}. Recursive division continues until a fixed size (threshold) that fits into processor caches.


\begin{figure}
  \centering
  \begin{lstlisting}[language=c++, basicstyle=\footnotesize]
 void recursiveMM(Matrix &C, const Matrix &A,
    const Matrix &B) {
  if (size <= threshold) {
    MatrixMultiplySimple(C, A, B);
    return;
  }
  recursiveMM(C.UL(), A.UL(), B.UL());
  recursiveMM(C.LL(), A.LL(), B.UL());
  recursiveMM(C.UR(), A.UL(), B.UR());
  recursiveMM(C.LR(), A.LL(), B.UR());
  recursiveMM(C.UL(), A.UR(), B.LL());
  recursiveMM(C.LL(), A.LR(), B.LL());
  recursiveMM(C.UR(), A.UR(), B.LR());
  recursiveMM(C.LR(), A.LR(), B.LR());
}

\end{lstlisting}
  \caption{Simple recursive matrix multiplication implementation. UL, UR, LL, and LR stand for the upper left, upper right, lower left and lower right sub-matrices.}
  \label{fig:mat-recursive}
\end{figure}

This formulation implements the $O(n^3)$ algorithm for $n \times n$ matrices. Larger matrices can benefit from Strassen's $O(n^{2.807})$ \cite{Strassen69} algorithm that avoids redundant computation at the expense of numerical stability. Strassen's uses a recursive reduction and can use MMM at the base of recursion. Algorithms with lower asymptotic complexity ($O(n^{2.371552})$~\cite{Williams23alpha}) are not used in practice. MMM does not implement Strassen's and is not optimized for larger matrices. Our effective performance range lies between $1024^2$ and $8192^2$.  However, Strassen's algorithm uses a similar recursive structure and requires these smaller block matrix multiplications.

Sparse Generalized Matrix Multiplication (SpGEMM)~\cite{spgemm} multiplies two matrices that are stored in a compressed format.
The goal is to have the computational work scale with the number of non-zeros in the matrix, rather than the matrix size.  SpGEMM stores the indexes and values of the non zeros compactly, this reduces computation, but loses some of the efficiency gained from aligned vector operations. For this reason, SpGEMM does not outperform dense dense matrix multiplication until high levels of sparsity. This is true when SpGEMM matrices are already encoded sparsely, i.e.~there is no conversion cost.
MMM will adapt some data structures from the SpGEMM literature.


\begin{figure}
  \centering
  \begin{tikzpicture}[element/.style={minimum width=.1cm,minimum height=0.1cm, draw}]
    \matrix (r) [matrix of nodes,nodes={element},column sep=-\pgflinewidth, row sep=-\pgflinewidth,]{
    |[fill=gray!00]|&|[fill=gray!00]|&|[fill=gray!00]|&|[fill=gray!00]|&|[fill=gray!00]|&|[fill=gray!60]|&|[fill=gray!60]|&|[fill=gray!00]|&|[fill=gray!00]|&|[fill=gray!00]|&|[fill=gray!00]|&|[fill=gray!00]|&|[fill=gray!00]|&|[fill=gray!00]|&|[fill=gray!00]|&|[fill=gray!00]|\\
    |[fill=gray!00]|&|[fill=gray!60]|&|[fill=gray!00]|&|[fill=gray!00]|&|[fill=gray!00]|&|[fill=gray!00]|&|[fill=gray!60]|&|[fill=gray!00]|&|[fill=gray!00]|&|[fill=gray!00]|&|[fill=gray!60]|&|[fill=gray!00]|&|[fill=gray!60]|&|[fill=gray!00]|&|[fill=gray!00]|&|[fill=gray!00]|\\
    |[fill=gray!60]|&|[fill=gray!60]|&|[fill=gray!60]|&|[fill=gray!60]|&|[fill=gray!00]|&|[fill=gray!00]|&|[fill=gray!00]|&|[fill=gray!00]|&|[fill=gray!00]|&|[fill=gray!60]|&|[fill=gray!00]|&|[fill=gray!00]|&|[fill=gray!60]|&|[fill=gray!00]|&|[fill=gray!00]|&|[fill=gray!00]|\\
    |[fill=gray!00]|&|[fill=gray!00]|&|[fill=gray!60]|&|[fill=gray!60]|&|[fill=gray!60]|&|[fill=gray!60]|&|[fill=gray!00]|&|[fill=gray!00]|&|[fill=gray!60]|&|[fill=gray!00]|&|[fill=gray!00]|&|[fill=gray!60]|&|[fill=gray!60]|&|[fill=gray!00]|&|[fill=gray!00]|&|[fill=gray!00]|\\
    |[fill=gray!60]|&|[fill=gray!60]|&|[fill=gray!60]|&|[fill=gray!00]|&|[fill=gray!00]|&|[fill=gray!00]|&|[fill=gray!00]|&|[fill=gray!00]|&|[fill=gray!00]|&|[fill=gray!00]|&|[fill=gray!00]|&|[fill=gray!00]|&|[fill=gray!60]|&|[fill=gray!00]|&|[fill=gray!60]|&|[fill=gray!00]|\\
    |[fill=gray!00]|&|[fill=gray!00]|&|[fill=gray!00]|&|[fill=gray!00]|&|[fill=gray!00]|&|[fill=gray!00]|&|[fill=gray!00]|&|[fill=gray!00]|&|[fill=gray!60]|&|[fill=gray!60]|&|[fill=gray!00]|&|[fill=gray!00]|&|[fill=gray!60]|&|[fill=gray!60]|&|[fill=gray!60]|&|[fill=gray!00]|\\
    |[fill=gray!00]|&|[fill=gray!60]|&|[fill=gray!60]|&|[fill=gray!60]|&|[fill=gray!60]|&|[fill=gray!00]|&|[fill=gray!00]|&|[fill=gray!00]|&|[fill=gray!60]|&|[fill=gray!60]|&|[fill=gray!00]|&|[fill=gray!00]|&|[fill=gray!00]|&|[fill=gray!60]|&|[fill=gray!00]|&|[fill=gray!00]|\\
    |[fill=gray!00]|&|[fill=gray!60]|&|[fill=gray!00]|&|[fill=gray!00]|&|[fill=gray!60]|&|[fill=gray!60]|&|[fill=gray!00]|&|[fill=gray!00]|&|[fill=gray!00]|&|[fill=gray!00]|&|[fill=gray!00]|&|[fill=gray!60]|&|[fill=gray!00]|&|[fill=gray!00]|&|[fill=gray!00]|&|[fill=gray!00]|\\
    |[fill=gray!60]|&|[fill=gray!60]|&|[fill=gray!00]|&|[fill=gray!00]|&|[fill=gray!00]|&|[fill=gray!00]|&|[fill=gray!00]|&|[fill=gray!00]|&|[fill=gray!00]|&|[fill=gray!00]|&|[fill=gray!00]|&|[fill=gray!00]|&|[fill=gray!00]|&|[fill=gray!00]|&|[fill=gray!00]|&|[fill=gray!60]|\\
    |[fill=gray!00]|&|[fill=gray!00]|&|[fill=gray!60]|&|[fill=gray!00]|&|[fill=gray!60]|&|[fill=gray!00]|&|[fill=gray!00]|&|[fill=gray!60]|&|[fill=gray!00]|&|[fill=gray!60]|&|[fill=gray!00]|&|[fill=gray!00]|&|[fill=gray!00]|&|[fill=gray!60]|&|[fill=gray!00]|&|[fill=gray!00]|\\
    |[fill=gray!00]|&|[fill=gray!60]|&|[fill=gray!00]|&|[fill=gray!00]|&|[fill=gray!00]|&|[fill=gray!00]|&|[fill=gray!00]|&|[fill=gray!00]|&|[fill=gray!60]|&|[fill=gray!00]|&|[fill=gray!00]|&|[fill=gray!60]|&|[fill=gray!00]|&|[fill=gray!00]|&|[fill=gray!00]|&|[fill=gray!00]|\\
    |[fill=gray!60]|&|[fill=gray!00]|&|[fill=gray!00]|&|[fill=gray!60]|&|[fill=gray!00]|&|[fill=gray!00]|&|[fill=gray!60]|&|[fill=gray!00]|&|[fill=gray!00]|&|[fill=gray!00]|&|[fill=gray!00]|&|[fill=gray!00]|&|[fill=gray!00]|&|[fill=gray!60]|&|[fill=gray!00]|&|[fill=gray!00]|\\
    |[fill=gray!00]|&|[fill=gray!60]|&|[fill=gray!60]|&|[fill=gray!00]|&|[fill=gray!60]|&|[fill=gray!60]|&|[fill=gray!00]|&|[fill=gray!00]|&|[fill=gray!00]|&|[fill=gray!00]|&|[fill=gray!00]|&|[fill=gray!00]|&|[fill=gray!00]|&|[fill=gray!60]|&|[fill=gray!00]|&|[fill=gray!00]|\\
    |[fill=gray!60]|&|[fill=gray!00]|&|[fill=gray!00]|&|[fill=gray!00]|&|[fill=gray!00]|&|[fill=gray!00]|&|[fill=gray!00]|&|[fill=gray!00]|&|[fill=gray!00]|&|[fill=gray!60]|&|[fill=gray!00]|&|[fill=gray!00]|&|[fill=gray!00]|&|[fill=gray!00]|&|[fill=gray!00]|&|[fill=gray!00]|\\
    |[fill=gray!60]|&|[fill=gray!60]|&|[fill=gray!00]|&|[fill=gray!60]|&|[fill=gray!00]|&|[fill=gray!00]|&|[fill=gray!00]|&|[fill=gray!00]|&|[fill=gray!60]|&|[fill=gray!00]|&|[fill=gray!00]|&|[fill=gray!00]|&|[fill=gray!60]|&|[fill=gray!00]|&|[fill=gray!00]|&|[fill=gray!00]|\\
    |[fill=gray!00]|&|[fill=gray!00]|&|[fill=gray!60]|&|[fill=gray!00]|&|[fill=gray!00]|&|[fill=gray!00]|&|[fill=gray!00]|&|[fill=gray!60]|&|[fill=gray!60]|&|[fill=gray!00]|&|[fill=gray!00]|&|[fill=gray!00]|&|[fill=gray!60]|&|[fill=gray!00]|&|[fill=gray!60]|&|[fill=gray!60]|\\
    };
    \node[above=0.1cm] at ($(r-1-8)!0.5!(r-1-9)$){\textbf{Random Sparsity}};

    \matrix (b) at (4.5,0) [matrix of nodes,nodes={element},column sep=-\pgflinewidth, row sep=-\pgflinewidth,]{
    |[fill=gray!00]|&|[fill=gray!00]|&|[fill=gray!00]|&|[fill=gray!00]|&|[fill=gray!60]|&|[fill=gray!60]|&|[fill=gray!60]|&|[fill=gray!60]|&|[fill=gray!60]|&|[fill=gray!60]|&|[fill=gray!60]|&|[fill=gray!60]|&|[fill=gray!00]|&|[fill=gray!00]|&|[fill=gray!00]|&|[fill=gray!00]|\\
    |[fill=gray!00]|&|[fill=gray!00]|&|[fill=gray!00]|&|[fill=gray!00]|&|[fill=gray!00]|&|[fill=gray!00]|&|[fill=gray!00]|&|[fill=gray!00]|&|[fill=gray!00]|&|[fill=gray!00]|&|[fill=gray!00]|&|[fill=gray!00]|&|[fill=gray!00]|&|[fill=gray!00]|&|[fill=gray!00]|&|[fill=gray!00]|\\
    |[fill=gray!60]|&|[fill=gray!60]|&|[fill=gray!60]|&|[fill=gray!60]|&|[fill=gray!00]|&|[fill=gray!00]|&|[fill=gray!00]|&|[fill=gray!00]|&|[fill=gray!00]|&|[fill=gray!00]|&|[fill=gray!00]|&|[fill=gray!00]|&|[fill=gray!00]|&|[fill=gray!00]|&|[fill=gray!00]|&|[fill=gray!00]|\\
    |[fill=gray!00]|&|[fill=gray!00]|&|[fill=gray!00]|&|[fill=gray!00]|&|[fill=gray!00]|&|[fill=gray!00]|&|[fill=gray!00]|&|[fill=gray!00]|&|[fill=gray!60]|&|[fill=gray!60]|&|[fill=gray!60]|&|[fill=gray!60]|&|[fill=gray!00]|&|[fill=gray!00]|&|[fill=gray!00]|&|[fill=gray!00]|\\
    |[fill=gray!00]|&|[fill=gray!00]|&|[fill=gray!00]|&|[fill=gray!00]|&|[fill=gray!00]|&|[fill=gray!00]|&|[fill=gray!00]|&|[fill=gray!00]|&|[fill=gray!00]|&|[fill=gray!00]|&|[fill=gray!00]|&|[fill=gray!00]|&|[fill=gray!00]|&|[fill=gray!00]|&|[fill=gray!00]|&|[fill=gray!00]|\\
    |[fill=gray!00]|&|[fill=gray!00]|&|[fill=gray!00]|&|[fill=gray!00]|&|[fill=gray!00]|&|[fill=gray!00]|&|[fill=gray!00]|&|[fill=gray!00]|&|[fill=gray!00]|&|[fill=gray!00]|&|[fill=gray!00]|&|[fill=gray!00]|&|[fill=gray!00]|&|[fill=gray!00]|&|[fill=gray!00]|&|[fill=gray!00]|\\
    |[fill=gray!60]|&|[fill=gray!60]|&|[fill=gray!60]|&|[fill=gray!60]|&|[fill=gray!00]|&|[fill=gray!00]|&|[fill=gray!00]|&|[fill=gray!00]|&|[fill=gray!60]|&|[fill=gray!60]|&|[fill=gray!60]|&|[fill=gray!60]|&|[fill=gray!00]|&|[fill=gray!00]|&|[fill=gray!00]|&|[fill=gray!00]|\\
    |[fill=gray!00]|&|[fill=gray!00]|&|[fill=gray!00]|&|[fill=gray!00]|&|[fill=gray!00]|&|[fill=gray!00]|&|[fill=gray!00]|&|[fill=gray!00]|&|[fill=gray!60]|&|[fill=gray!60]|&|[fill=gray!60]|&|[fill=gray!60]|&|[fill=gray!60]|&|[fill=gray!60]|&|[fill=gray!60]|&|[fill=gray!60]|\\
    |[fill=gray!00]|&|[fill=gray!00]|&|[fill=gray!00]|&|[fill=gray!00]|&|[fill=gray!00]|&|[fill=gray!00]|&|[fill=gray!00]|&|[fill=gray!00]|&|[fill=gray!00]|&|[fill=gray!00]|&|[fill=gray!00]|&|[fill=gray!00]|&|[fill=gray!00]|&|[fill=gray!00]|&|[fill=gray!00]|&|[fill=gray!00]|\\
    |[fill=gray!00]|&|[fill=gray!00]|&|[fill=gray!00]|&|[fill=gray!00]|&|[fill=gray!00]|&|[fill=gray!00]|&|[fill=gray!00]|&|[fill=gray!00]|&|[fill=gray!00]|&|[fill=gray!00]|&|[fill=gray!00]|&|[fill=gray!00]|&|[fill=gray!00]|&|[fill=gray!00]|&|[fill=gray!00]|&|[fill=gray!00]|\\
    |[fill=gray!00]|&|[fill=gray!00]|&|[fill=gray!00]|&|[fill=gray!00]|&|[fill=gray!00]|&|[fill=gray!00]|&|[fill=gray!00]|&|[fill=gray!00]|&|[fill=gray!60]|&|[fill=gray!60]|&|[fill=gray!60]|&|[fill=gray!60]|&|[fill=gray!00]|&|[fill=gray!00]|&|[fill=gray!00]|&|[fill=gray!00]|\\
    |[fill=gray!00]|&|[fill=gray!00]|&|[fill=gray!00]|&|[fill=gray!00]|&|[fill=gray!00]|&|[fill=gray!00]|&|[fill=gray!00]|&|[fill=gray!00]|&|[fill=gray!00]|&|[fill=gray!00]|&|[fill=gray!00]|&|[fill=gray!00]|&|[fill=gray!60]|&|[fill=gray!60]|&|[fill=gray!60]|&|[fill=gray!60]|\\
    |[fill=gray!60]|&|[fill=gray!60]|&|[fill=gray!60]|&|[fill=gray!60]|&|[fill=gray!00]|&|[fill=gray!00]|&|[fill=gray!00]|&|[fill=gray!00]|&|[fill=gray!00]|&|[fill=gray!00]|&|[fill=gray!00]|&|[fill=gray!00]|&|[fill=gray!00]|&|[fill=gray!00]|&|[fill=gray!00]|&|[fill=gray!00]|\\
    |[fill=gray!00]|&|[fill=gray!00]|&|[fill=gray!00]|&|[fill=gray!00]|&|[fill=gray!00]|&|[fill=gray!00]|&|[fill=gray!00]|&|[fill=gray!00]|&|[fill=gray!00]|&|[fill=gray!00]|&|[fill=gray!00]|&|[fill=gray!00]|&|[fill=gray!00]|&|[fill=gray!00]|&|[fill=gray!00]|&|[fill=gray!00]|\\
    |[fill=gray!60]|&|[fill=gray!60]|&|[fill=gray!60]|&|[fill=gray!60]|&|[fill=gray!00]|&|[fill=gray!00]|&|[fill=gray!00]|&|[fill=gray!00]|&|[fill=gray!00]|&|[fill=gray!00]|&|[fill=gray!00]|&|[fill=gray!00]|&|[fill=gray!00]|&|[fill=gray!00]|&|[fill=gray!00]|&|[fill=gray!00]|\\
    |[fill=gray!00]|&|[fill=gray!00]|&|[fill=gray!00]|&|[fill=gray!00]|&|[fill=gray!00]|&|[fill=gray!00]|&|[fill=gray!00]|&|[fill=gray!00]|&|[fill=gray!00]|&|[fill=gray!00]|&|[fill=gray!00]|&|[fill=gray!00]|&|[fill=gray!60]|&|[fill=gray!60]|&|[fill=gray!60]|&|[fill=gray!60]|\\
    };
    \node[above=0.1cm] at ($(b-1-8)!0.5!(b-1-9)$){\textbf{Block Random Sparsity}};

    \matrix (c) at (0,-4) [matrix of nodes,nodes={element},column sep=-\pgflinewidth, row sep=-\pgflinewidth,]{
    |[fill=gray!00]|&|[fill=gray!60]|&|[fill=gray!00]|&|[fill=gray!00]|&|[fill=gray!60]|&|[fill=gray!00]|&|[fill=gray!00]|&|[fill=gray!00]|&|[fill=gray!00]|&|[fill=gray!00]|&|[fill=gray!00]|&|[fill=gray!60]|&|[fill=gray!00]|&|[fill=gray!60]|&|[fill=gray!00]|&|[fill=gray!00]|\\
    |[fill=gray!00]|&|[fill=gray!60]|&|[fill=gray!00]|&|[fill=gray!00]|&|[fill=gray!60]|&|[fill=gray!00]|&|[fill=gray!00]|&|[fill=gray!00]|&|[fill=gray!00]|&|[fill=gray!00]|&|[fill=gray!00]|&|[fill=gray!60]|&|[fill=gray!00]|&|[fill=gray!60]|&|[fill=gray!00]|&|[fill=gray!00]|\\
    |[fill=gray!00]|&|[fill=gray!60]|&|[fill=gray!00]|&|[fill=gray!00]|&|[fill=gray!60]|&|[fill=gray!00]|&|[fill=gray!00]|&|[fill=gray!00]|&|[fill=gray!00]|&|[fill=gray!00]|&|[fill=gray!00]|&|[fill=gray!60]|&|[fill=gray!00]|&|[fill=gray!60]|&|[fill=gray!00]|&|[fill=gray!00]|\\
    |[fill=gray!00]|&|[fill=gray!60]|&|[fill=gray!00]|&|[fill=gray!00]|&|[fill=gray!60]|&|[fill=gray!00]|&|[fill=gray!00]|&|[fill=gray!00]|&|[fill=gray!00]|&|[fill=gray!00]|&|[fill=gray!00]|&|[fill=gray!60]|&|[fill=gray!00]|&|[fill=gray!60]|&|[fill=gray!00]|&|[fill=gray!00]|\\
    |[fill=gray!00]|&|[fill=gray!60]|&|[fill=gray!00]|&|[fill=gray!00]|&|[fill=gray!60]|&|[fill=gray!00]|&|[fill=gray!00]|&|[fill=gray!00]|&|[fill=gray!00]|&|[fill=gray!00]|&|[fill=gray!00]|&|[fill=gray!60]|&|[fill=gray!00]|&|[fill=gray!60]|&|[fill=gray!00]|&|[fill=gray!00]|\\
    |[fill=gray!00]|&|[fill=gray!60]|&|[fill=gray!00]|&|[fill=gray!00]|&|[fill=gray!60]|&|[fill=gray!00]|&|[fill=gray!00]|&|[fill=gray!00]|&|[fill=gray!00]|&|[fill=gray!00]|&|[fill=gray!00]|&|[fill=gray!60]|&|[fill=gray!00]|&|[fill=gray!60]|&|[fill=gray!00]|&|[fill=gray!00]|\\
    |[fill=gray!00]|&|[fill=gray!60]|&|[fill=gray!00]|&|[fill=gray!00]|&|[fill=gray!60]|&|[fill=gray!00]|&|[fill=gray!00]|&|[fill=gray!00]|&|[fill=gray!00]|&|[fill=gray!00]|&|[fill=gray!00]|&|[fill=gray!60]|&|[fill=gray!00]|&|[fill=gray!60]|&|[fill=gray!00]|&|[fill=gray!00]|\\
    |[fill=gray!00]|&|[fill=gray!60]|&|[fill=gray!00]|&|[fill=gray!00]|&|[fill=gray!60]|&|[fill=gray!00]|&|[fill=gray!00]|&|[fill=gray!00]|&|[fill=gray!00]|&|[fill=gray!00]|&|[fill=gray!00]|&|[fill=gray!60]|&|[fill=gray!00]|&|[fill=gray!60]|&|[fill=gray!00]|&|[fill=gray!00]|\\
    |[fill=gray!00]|&|[fill=gray!60]|&|[fill=gray!00]|&|[fill=gray!00]|&|[fill=gray!60]|&|[fill=gray!00]|&|[fill=gray!00]|&|[fill=gray!00]|&|[fill=gray!00]|&|[fill=gray!00]|&|[fill=gray!00]|&|[fill=gray!60]|&|[fill=gray!00]|&|[fill=gray!60]|&|[fill=gray!00]|&|[fill=gray!00]|\\
    |[fill=gray!00]|&|[fill=gray!60]|&|[fill=gray!00]|&|[fill=gray!00]|&|[fill=gray!60]|&|[fill=gray!00]|&|[fill=gray!00]|&|[fill=gray!00]|&|[fill=gray!00]|&|[fill=gray!00]|&|[fill=gray!00]|&|[fill=gray!60]|&|[fill=gray!00]|&|[fill=gray!60]|&|[fill=gray!00]|&|[fill=gray!00]|\\
    |[fill=gray!00]|&|[fill=gray!60]|&|[fill=gray!00]|&|[fill=gray!00]|&|[fill=gray!60]|&|[fill=gray!00]|&|[fill=gray!00]|&|[fill=gray!00]|&|[fill=gray!00]|&|[fill=gray!00]|&|[fill=gray!00]|&|[fill=gray!60]|&|[fill=gray!00]|&|[fill=gray!60]|&|[fill=gray!00]|&|[fill=gray!00]|\\
    |[fill=gray!00]|&|[fill=gray!60]|&|[fill=gray!00]|&|[fill=gray!00]|&|[fill=gray!60]|&|[fill=gray!00]|&|[fill=gray!00]|&|[fill=gray!00]|&|[fill=gray!00]|&|[fill=gray!00]|&|[fill=gray!00]|&|[fill=gray!60]|&|[fill=gray!00]|&|[fill=gray!60]|&|[fill=gray!00]|&|[fill=gray!00]|\\
    |[fill=gray!00]|&|[fill=gray!60]|&|[fill=gray!00]|&|[fill=gray!00]|&|[fill=gray!60]|&|[fill=gray!00]|&|[fill=gray!00]|&|[fill=gray!00]|&|[fill=gray!00]|&|[fill=gray!00]|&|[fill=gray!00]|&|[fill=gray!60]|&|[fill=gray!00]|&|[fill=gray!60]|&|[fill=gray!00]|&|[fill=gray!00]|\\
    |[fill=gray!00]|&|[fill=gray!60]|&|[fill=gray!00]|&|[fill=gray!00]|&|[fill=gray!60]|&|[fill=gray!00]|&|[fill=gray!00]|&|[fill=gray!00]|&|[fill=gray!00]|&|[fill=gray!00]|&|[fill=gray!00]|&|[fill=gray!60]|&|[fill=gray!00]|&|[fill=gray!60]|&|[fill=gray!00]|&|[fill=gray!00]|\\
    |[fill=gray!00]|&|[fill=gray!60]|&|[fill=gray!00]|&|[fill=gray!00]|&|[fill=gray!60]|&|[fill=gray!00]|&|[fill=gray!00]|&|[fill=gray!00]|&|[fill=gray!00]|&|[fill=gray!00]|&|[fill=gray!00]|&|[fill=gray!60]|&|[fill=gray!00]|&|[fill=gray!60]|&|[fill=gray!00]|&|[fill=gray!00]|\\
    |[fill=gray!00]|&|[fill=gray!60]|&|[fill=gray!00]|&|[fill=gray!00]|&|[fill=gray!60]|&|[fill=gray!00]|&|[fill=gray!00]|&|[fill=gray!00]|&|[fill=gray!00]|&|[fill=gray!00]|&|[fill=gray!00]|&|[fill=gray!60]|&|[fill=gray!00]|&|[fill=gray!60]|&|[fill=gray!00]|&|[fill=gray!00]|\\
    };
    \node[above=0.1cm] at ($(c-1-8)!0.5!(c-1-9)$){\textbf{Random Column Sparsity}};

    \matrix (bc) at (4.5,-4) [matrix of nodes,nodes={element},column sep=-\pgflinewidth, row sep=-\pgflinewidth,]{
    |[fill=gray!00]|&|[fill=gray!00]|&|[fill=gray!00]|&|[fill=gray!00]|&|[fill=gray!00]|&|[fill=gray!00]|&|[fill=gray!00]|&|[fill=gray!00]|&|[fill=gray!60]|&|[fill=gray!60]|&|[fill=gray!60]|&|[fill=gray!60]|&|[fill=gray!00]|&|[fill=gray!00]|&|[fill=gray!00]|&|[fill=gray!00]|\\
    |[fill=gray!00]|&|[fill=gray!00]|&|[fill=gray!00]|&|[fill=gray!00]|&|[fill=gray!00]|&|[fill=gray!00]|&|[fill=gray!00]|&|[fill=gray!00]|&|[fill=gray!60]|&|[fill=gray!60]|&|[fill=gray!60]|&|[fill=gray!60]|&|[fill=gray!00]|&|[fill=gray!00]|&|[fill=gray!00]|&|[fill=gray!00]|\\
    |[fill=gray!00]|&|[fill=gray!00]|&|[fill=gray!00]|&|[fill=gray!00]|&|[fill=gray!00]|&|[fill=gray!00]|&|[fill=gray!00]|&|[fill=gray!00]|&|[fill=gray!60]|&|[fill=gray!60]|&|[fill=gray!60]|&|[fill=gray!60]|&|[fill=gray!00]|&|[fill=gray!00]|&|[fill=gray!00]|&|[fill=gray!00]|\\
    |[fill=gray!00]|&|[fill=gray!00]|&|[fill=gray!00]|&|[fill=gray!00]|&|[fill=gray!00]|&|[fill=gray!00]|&|[fill=gray!00]|&|[fill=gray!00]|&|[fill=gray!60]|&|[fill=gray!60]|&|[fill=gray!60]|&|[fill=gray!60]|&|[fill=gray!00]|&|[fill=gray!00]|&|[fill=gray!00]|&|[fill=gray!00]|\\
    |[fill=gray!00]|&|[fill=gray!00]|&|[fill=gray!00]|&|[fill=gray!00]|&|[fill=gray!00]|&|[fill=gray!00]|&|[fill=gray!00]|&|[fill=gray!00]|&|[fill=gray!60]|&|[fill=gray!60]|&|[fill=gray!60]|&|[fill=gray!60]|&|[fill=gray!00]|&|[fill=gray!00]|&|[fill=gray!00]|&|[fill=gray!00]|\\
    |[fill=gray!00]|&|[fill=gray!00]|&|[fill=gray!00]|&|[fill=gray!00]|&|[fill=gray!00]|&|[fill=gray!00]|&|[fill=gray!00]|&|[fill=gray!00]|&|[fill=gray!60]|&|[fill=gray!60]|&|[fill=gray!60]|&|[fill=gray!60]|&|[fill=gray!00]|&|[fill=gray!00]|&|[fill=gray!00]|&|[fill=gray!00]|\\
    |[fill=gray!00]|&|[fill=gray!00]|&|[fill=gray!00]|&|[fill=gray!00]|&|[fill=gray!00]|&|[fill=gray!00]|&|[fill=gray!00]|&|[fill=gray!00]|&|[fill=gray!60]|&|[fill=gray!60]|&|[fill=gray!60]|&|[fill=gray!60]|&|[fill=gray!00]|&|[fill=gray!00]|&|[fill=gray!00]|&|[fill=gray!00]|\\
    |[fill=gray!00]|&|[fill=gray!00]|&|[fill=gray!00]|&|[fill=gray!00]|&|[fill=gray!00]|&|[fill=gray!00]|&|[fill=gray!00]|&|[fill=gray!00]|&|[fill=gray!60]|&|[fill=gray!60]|&|[fill=gray!60]|&|[fill=gray!60]|&|[fill=gray!00]|&|[fill=gray!00]|&|[fill=gray!00]|&|[fill=gray!00]|\\
    |[fill=gray!00]|&|[fill=gray!00]|&|[fill=gray!00]|&|[fill=gray!00]|&|[fill=gray!00]|&|[fill=gray!00]|&|[fill=gray!00]|&|[fill=gray!00]|&|[fill=gray!60]|&|[fill=gray!60]|&|[fill=gray!60]|&|[fill=gray!60]|&|[fill=gray!00]|&|[fill=gray!00]|&|[fill=gray!00]|&|[fill=gray!00]|\\
    |[fill=gray!00]|&|[fill=gray!00]|&|[fill=gray!00]|&|[fill=gray!00]|&|[fill=gray!00]|&|[fill=gray!00]|&|[fill=gray!00]|&|[fill=gray!00]|&|[fill=gray!60]|&|[fill=gray!60]|&|[fill=gray!60]|&|[fill=gray!60]|&|[fill=gray!00]|&|[fill=gray!00]|&|[fill=gray!00]|&|[fill=gray!00]|\\
    |[fill=gray!00]|&|[fill=gray!00]|&|[fill=gray!00]|&|[fill=gray!00]|&|[fill=gray!00]|&|[fill=gray!00]|&|[fill=gray!00]|&|[fill=gray!00]|&|[fill=gray!60]|&|[fill=gray!60]|&|[fill=gray!60]|&|[fill=gray!60]|&|[fill=gray!00]|&|[fill=gray!00]|&|[fill=gray!00]|&|[fill=gray!00]|\\
    |[fill=gray!00]|&|[fill=gray!00]|&|[fill=gray!00]|&|[fill=gray!00]|&|[fill=gray!00]|&|[fill=gray!00]|&|[fill=gray!00]|&|[fill=gray!00]|&|[fill=gray!60]|&|[fill=gray!60]|&|[fill=gray!60]|&|[fill=gray!60]|&|[fill=gray!00]|&|[fill=gray!00]|&|[fill=gray!00]|&|[fill=gray!00]|\\
    |[fill=gray!00]|&|[fill=gray!00]|&|[fill=gray!00]|&|[fill=gray!00]|&|[fill=gray!00]|&|[fill=gray!00]|&|[fill=gray!00]|&|[fill=gray!00]|&|[fill=gray!60]|&|[fill=gray!60]|&|[fill=gray!60]|&|[fill=gray!60]|&|[fill=gray!00]|&|[fill=gray!00]|&|[fill=gray!00]|&|[fill=gray!00]|\\
    |[fill=gray!00]|&|[fill=gray!00]|&|[fill=gray!00]|&|[fill=gray!00]|&|[fill=gray!00]|&|[fill=gray!00]|&|[fill=gray!00]|&|[fill=gray!00]|&|[fill=gray!60]|&|[fill=gray!60]|&|[fill=gray!60]|&|[fill=gray!60]|&|[fill=gray!00]|&|[fill=gray!00]|&|[fill=gray!00]|&|[fill=gray!00]|\\
    |[fill=gray!00]|&|[fill=gray!00]|&|[fill=gray!00]|&|[fill=gray!00]|&|[fill=gray!00]|&|[fill=gray!00]|&|[fill=gray!00]|&|[fill=gray!00]|&|[fill=gray!60]|&|[fill=gray!60]|&|[fill=gray!60]|&|[fill=gray!60]|&|[fill=gray!00]|&|[fill=gray!00]|&|[fill=gray!00]|&|[fill=gray!00]|\\
    |[fill=gray!00]|&|[fill=gray!00]|&|[fill=gray!00]|&|[fill=gray!00]|&|[fill=gray!00]|&|[fill=gray!00]|&|[fill=gray!00]|&|[fill=gray!00]|&|[fill=gray!60]|&|[fill=gray!60]|&|[fill=gray!60]|&|[fill=gray!60]|&|[fill=gray!00]|&|[fill=gray!00]|&|[fill=gray!00]|&|[fill=gray!00]|\\
    };
    \node[above=0.1cm] at ($(bc-1-8)!0.5!(bc-1-9)$){\textbf{Block Random Column Sparsity}};

  \end{tikzpicture}
  \caption{Types of sparsity. In random sparsity, any bit may be non-zero. Block-random sparsity assumes aligned, contiguous regions of each row are sparse or not sparse. The degree of block-random sparsity may be lower than the fraction of sparsity when 0s are not fully aligned. We also show random column and random block column sparsity. Block column patterns arise in Deja Vu~\cite{liu2023deja}.}
  \label{fig:matrix_forms}
\end{figure}
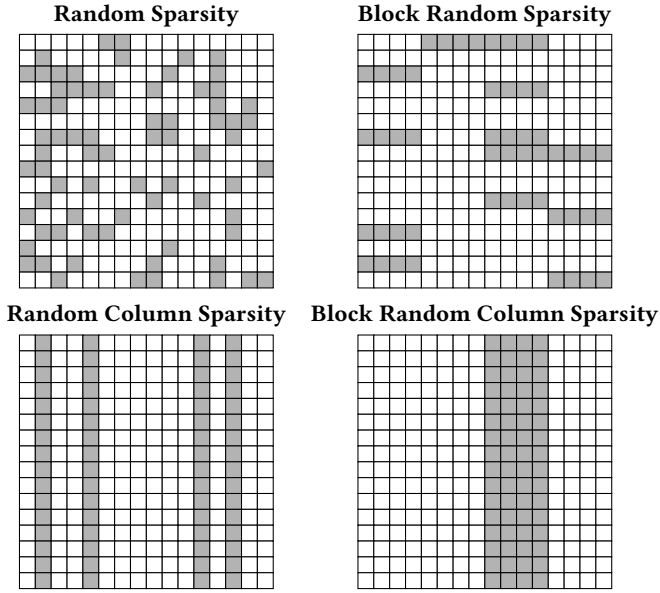

\section{Masked Matrix Multiplication}
\label{sec:MMM}

We provide improvements to matrix multiplication for emergent and contextual sparsity by detecting zero-values at runtime to avoid unnecessary computations. The challenge lies in detecting zero values efficiently while preserving the multicore and vector parallelism of the underlying hardware. The implementation sits in between dense matrix multiplication and SpGEMM. Dense matrix multiplication maximizes memory throughput and hardware parallelism, but computes all elements. SpGEMM only computes non-zero elements, but uses data structures that require branching instructions and perform more random memory accesses.

We call this Masked Matrix Multiplication (MMM) because the sparsity arises at runtime. The underlying data are dense so that sparsity cannot be evaluated at compile time and the data cannot be stored in sparse formats. We express this as $M(X) \times N(Y) = C$ in which $M$ and $N$ are masking functions over dense arrays $X$ and $Y$, we define $A=M(X)$ and $B=N(Y)$. Performance benefits arise when both $A$ and $B$ are sparse; if only one is sparse, simpler techniques are more efficient. More specifically, we require random bit sparsity in the $A$ matrix evaluated in the outer loop and block random sparsity in $B$ (Figure \ref{fig:matrix_forms}).

\vspace{5pt}

\begin{figure}
  \centering
  \begin{lstlisting}[language=c++, basicstyle=\footnotesize]
 for (int i = 0; i < C.rows; ++i) {
   for (int k = 0; k < B.rows; ++k) {
     auto A_val = A[i, k];
     if (A_val == 0) {
       continue;
     }
     for (int j = 0; j < C.cols; ++j) {
       C[i, j] += A_val * B[k, j];
     }
   }
 }


\end{lstlisting}
  \caption{A simple optimization to skip computation when A is sparse.  The {\tt if} statement skips over rows of B when the corresponding element of A is 0.}
  \label{fig:mask_A}
\end{figure}

\noindent \textbf{Random Sparsity in $A$.} We evaluate elements of $A$ one at a time in our middle loop and avoid computation when elements are zero. This is a standard technique. It is customary when choosing loop orders for tensors operations~\cite{taco} or in hardware implementations of sparse accelerators~\cite{wu2021sparseloop} to lift the sparsest matrix to the outer loop for exactly this reason. A branch instruction,
{\tt if (A\_val == 0)}, avoids evaluation of the inner loop (Figure \ref{fig:mask_A}).


\vspace{5pt}

\noindent This skips an entire row of $B$ and an entire row of partial writes to $C$. In later steps, we will optimize the evaluation
of the branch instruction.

\vspace{5pt}

\noindent \textbf{Block-Random Sparsity in $B$}. Eliminating computation in the inner loop is more complicated, because efficient implementations preserve vector processing of multiple elements. We cannot eliminate computation on a single 0 element.
The compiler turns the inner loop into a vector computation for which the psuedocode can be seen in Figure \ref{fig:vector_base}.
The inner operation is a vector fused multiply add (FMA) and most architectures support this instruction natively,
including our targets Intel AVX-2 and AVX-512. There are masked variants of FMA that skip elements in the vector. However, masked operations have the same cost as FMA on the entire vector.

\begin{figure}
  \centering
  \begin{lstlisting}[language=c++, basicstyle=\footnotesize]
for (index_t j = 0; j < C.cols; j += vector_size) {
  C[i, j:j+vector_size] += 
    {A_val,...,A_val} * B[k, j:j+vector_size];
}

\end{lstlisting}
  \caption{The base vectorized row code to help compute matrix multiplication.}
  \label{fig:vector_base}
\end{figure}

\begin{figure*}
  \begin{center}
    \includegraphics[width=1.95\columnwidth]{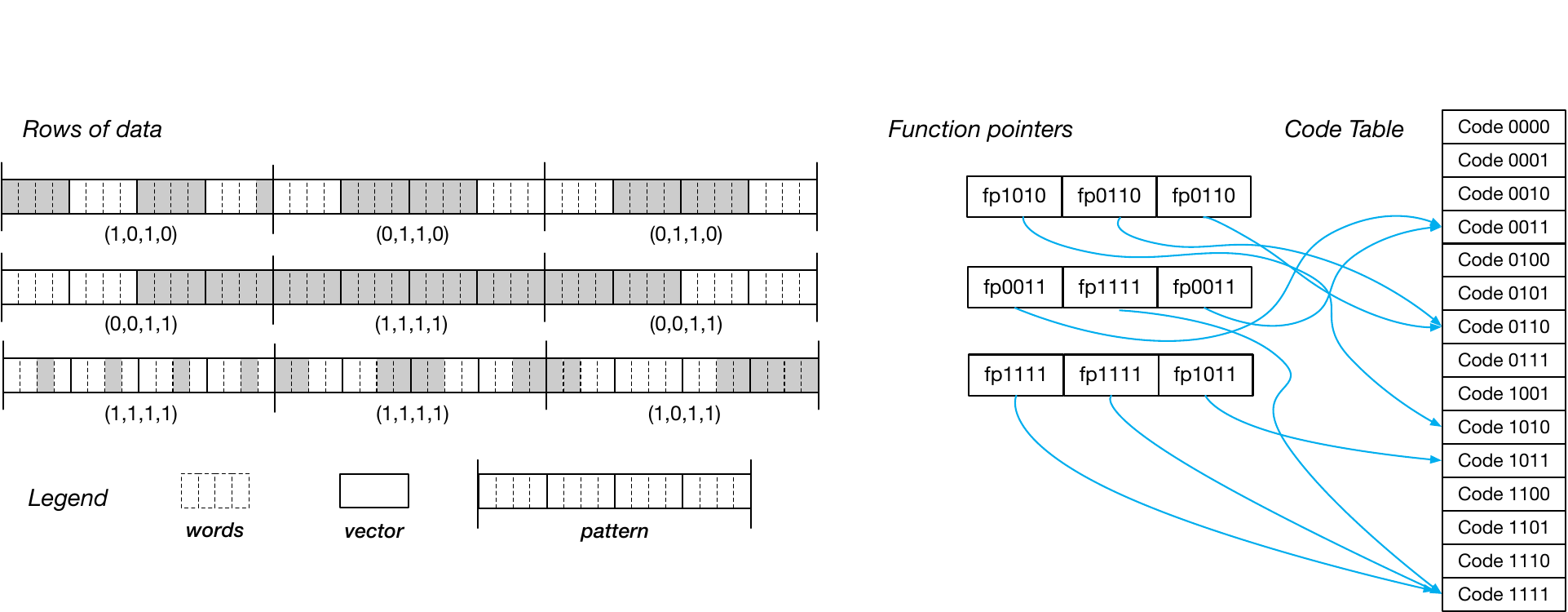}
    \caption{Preprocessing patterns of length 4 in $B$. Grayed entries are non-zero. A compare of each vector extracts a 0/1 bit indicating whether any words in the vector are non-zero. A pattern worth of bits is used to lookup a function pointer which is stored in an array or list of functions pointers. The function pointer array is used during multiplication to directly reference code that avoids computation on 0 vectors.}
    \label{fig:preprocess_B}
  \end{center}
  \vspace{5pt}
\end{figure*}


The goal is to identify entire vectors of 0s on which can skip computation. A simple approach checks that elements
$B[k, j:j+vector\_size]$ are all zero prior to performing the vector FMA. Unfortunately, checking that the vector is
zeros costs about as many cycles as the FMA. So, no savings are possible.




We will leverage that the rows of $B$ are used $C.cols$ times to preprocess {\em patterns} of sparsity, paying a one-time cost to encode sparsity and then incur only one memory reference of overhead per pattern to avoid computing on vectors of all zeros.
A pattern describes whether each vector in a fixed-width set of vectors within a row are all zeros.
At compile time, we generate optimized code for all possible patterns that we store in a code table.

Figure \ref{fig:preprocess_B} gives a schematic overview of the computation. Patterns consist of multiple vectors. We preprocess the rows of the $B$ matrix to extract patterns that we encode as bit strings. In this example, patterns of 4 in which $0$ indicates all zero values and $1$ indicates some non-zero values. Using the bit string, we build an array of function pointers to the optimized code for each pattern. During matrix multiplication, we will iterate over the $B$ matrix and the function pointers in lock step to select the optimized code for each pattern.


%

\vspace{5pt}

\noindent \textbf{Compile-Time Code Generation:} We build a code table for all possible vector patterns in which the instructions compute the non-zero vectors and skip over the zero vectors.
There are no branch instructions. Figure \ref{fig:row_code} gives the full set of AVX-2 256-bit vector instructions for patterns of size 8 on 32-bit floats. Each specific subversion takes the lines A-H corresponding to the non-zero pattern bits, e.g.~pattern 01010000 would take lines BDBDBD which would conduct load, fused multiply add, and store on only the second and fourth vector in the pattern.
All patterns are listed in a code table and we build a map of function pointers for each pattern. Code is not of fixed size. With AVX2, functions are as small as 1 byte (the empty case 00000000) to 166 bytes (the complete case 1111111) for a total size of 21376 bytes.

Choosing the size of the pattern has tradeoffs that affect compile time, instruction cache performance, and the granularity at which we can detect sparsity.  The number of patterns grows as $2^{size}$. Our implementation has 8 bits which results in 256 patterns. Large patterns have larger code tables, take more compile time, and perform fewer memory accesses. Smaller patterns tend to fit into L1i cache so that the memory indirection is less costly. We found that a pattern size of 8 works well on recent generations of Intel processors. Generating the code table takes a few seconds and performance nearly matches that when using 16 bit patterns\footnote{Generating the code table for a pattern size of 16 can take more than an hour and starts running into implementation limits for some compilers.}.






\begin{figure}

  \centering
  \begin{lstlisting}[language={[x86masm]Assembler}, basicstyle=\footnotesize]
A:   vmovups ymm1, ymmword ptr [rsi]
B:   vmovups ymm2, ymmword ptr [rsi + 32]
C:   vmovups ymm3, ymmword ptr [rsi + 64]
D:   vmovups ymm4, ymmword ptr [rsi + 96]
E:   vmovups ymm5, ymmword ptr [rsi + 128]
F:   vmovups ymm6, ymmword ptr [rsi + 160]
G:   vmovups ymm7, ymmword ptr [rsi + 192]
H:   vmovups ymm8, ymmword ptr [rsi + 224]

A:   vfmadd213ps     ymm1, ymm0, ymmword ptr rdi]
B:   vfmadd213ps     ymm2, ymm0, ymmword ptr rdi + 32]
C:   vfmadd213ps     ymm3, ymm0, ymmword ptr rdi + 64]
D:   vfmadd213ps     ymm4, ymm0, ymmword ptr rdi + 96]
E:   vfmadd213ps     ymm5, ymm0, ymmword ptr rdi + 128]
F:   vfmadd213ps     ymm6, ymm0, ymmword ptr rdi + 160]
G:   vfmadd213ps     ymm7, ymm0, ymmword ptr rdi + 192]
H:   vfmadd213ps     ymm8, ymm0, ymmword ptr [rdi + 224]

A:   vmovups ymmword ptr [rdi], ymm1
B:   vmovups ymmword ptr [rdi + 32], ymm2
C:   vmovups ymmword ptr [rdi + 64], ymm3
D:   vmovups ymmword ptr [rdi + 96], ymm4
E:   vmovups ymmword ptr [rdi + 128], ymm5
F:   vmovups ymmword ptr [rdi + 160], ymm6
G:   vmovups ymmword ptr [rdi + 192], ymm7
H:   vmovups ymmword ptr [rdi + 224], ymm8
    ret

\end{lstlisting}
  \caption{Full pattern multiplication code. Specific patterns will use a subset of lines A-H.
    At start ymm0 has the element from A in all positions, rsi is the address of the row of B, rdi is the address to the row of C.}
  \label{fig:row_code}
\end{figure}

\vspace{5pt}

\noindent\textbf{Preprocessing}: We take a single pass over each sub-matrix of the matrix evaluating sparsity for each pattern and build an array of function pointers to the optimized code. A sub-matrix is the unit and the bottom of the recursion that actually computes the matrix multiplication as seen in Figure \ref{fig:mat-recursive}.

This process is efficient, because it accesses memory sequentially, comparisons are vector aligned, and it writes function pointers sequentially. We preprocess sub-matrices of  B in the same order that we multiply in the next phase so that function pointers are stored in the order that they will be accessed.  The preprocessing pseudocode (Figure \ref{fig:fps}) looks at a pattern at a time (outer for loop) and does a compare to determine if each vector is all 0s (inner for loop) to extract a bit for each vector. The {\tt pattern\_index} bitstring is used to lookup the function pointer and write it to the {\sf fps} function pointer array. This computation is fully aligned. This is in contrast to the computation needed to convert a dense matrix to a sparse matrix format \cite{bulucc2012parallel} that requires packing elements together. Conversion to sparse formats is difficult to vectorize.

This preprocessing code is called a single time for each sub-matrix at a cost of $O(n^2)$ for $n \times n$ dense matrices.  The cost of preprocessing becomes negligible as the matrix size grows.  Empirically we find that this crossover point to be around $1024^2$.  This code can also be trivially parallelized over the different sub-matrices.

\begin{figure}

  \centering
  \begin{lstlisting}[language=python, basicstyle=\footnotesize]
# function pointers for sub-matrix
fps = []
for r in range(rows):
  for c in range(0, cols, vector_size*pattern_size):
    pattern = 0
    pattern_index = 1 << pattern_size
    for p_index in range(c, 
        c+vector_size*pattern_size, vector_size):
      # vector comparison
      if submatrix(r, p_index:p_index+vector_size) == 0:
          pattern |= pattern_index
      pattern_index >>= 1
    fps[submatrix_index].append(
        function_pointer_table[pattern])
\end{lstlisting}
  \caption{Preprocessing function pointer arrays that encode pattern sparsity. All computations are vector aligned.}
  \label{fig:fps}
\end{figure}

\vspace{5pt}

\noindent\textbf{Matrix Multiplication} (Figure \ref{fig:MMM}) loops over the elements by regions of size {\tt vector\_size * pattern\_size} and calls the corresponding optimized function {\tt fps[fp\_index]}.  This pays a single indirect call for each region. The optimized functions skip over all zero vectors using no branch instructions.

\begin{figure}

  \centering
  \begin{lstlisting}[language=python, basicstyle=\footnotesize]
step_size = vector_size * pattern_size
for j in range(A.rows):
  fp_index = 0
  for k in range(B.rows):
    if A[i, k] == 0:
        # skip region of fps corrosponding to the zero element of A
        fp_index += B.cols / step_size
        continue
    for j in range(0, B.cols, step_size):
      fps[fp_index](&C[i,j], A[i,k], &B[k,j])
      fp_index+=1

\end{lstlisting}
  \caption{MMM Pseudocode.  The MMM computation has the same iteration structure as the dense matrix multiply code, but skips zeros with the optimized row functions.}
  \label{fig:MMM}
\end{figure}

\subsection*{Optimizations}

\begin{figure}
  \begin{center}
    \includegraphics[width=0.9\columnwidth]{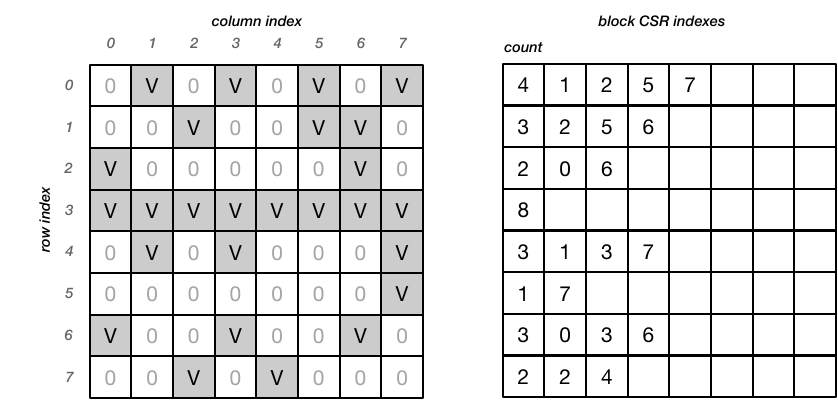}
    \caption{Blocked sparse row preprocessing of sub-matrices of $A$. In the input (left), grayed entries are $V$ indicate a value at a location. The output (right) encodes the count of non-zero elements and their offsets. The offsets in a full row (count of 8) do not need to be stored explicitly.}
    \label{fig:block_csr}
  \end{center}
\end{figure}

Our first optimization helps avoid branch instructions as $A$ gets more sparse. We preprocess $A$ on $256^2$ blocks to encode sparse indexes (Figure \ref{fig:block_csr}). We build a fully aligned data structure of the same size that encodes the count of non-zero elements and their offsets. This allows us to replace the branch {\tt if A[i, k] == 0:} with a for loop over the-nonzero elements so that the number of conditional branches scales with the number of non-zeros, rather than with the number of columns. This is a memory aligned version of the compressed sparse row format~\cite{bulucc2012parallel}.


Another optimization increases the size of the base vector on which we operate. It is possible to have each pattern represent more than 1 vector, e.g. 2 or 4 vectors. Doubling the base vector size halves the number of calls to the optimized sub-matrix computation, but also doubles the size of the smallest block pattern we capture. We see benefits here for some workloads.

\section{Evaluation}

We explore the potential benefits of MMM in comparison with dense-dense, sparse-dense, and sparse-sparse multiplication in Intel's MKL library. There is an extremely complicated performance landscape among matrix size, variations is sparsity in A and B matrices, parallelism, and architecture. Experiments against synthetic data on a consumer grade machine vary these parameters to characterize this landscape and show the regions in which MMM outperforms all MKL techniques. These will be mid-ranges of sparsity, mid-size matrices, and multi-core parallelism. We extend these results to server-class machines and on data sets extracted from Deja Vu~\cite{liu2023deja}. Experiments show improvements on select cases. We do not implement hardware optimizations (register reuse, pipelining) beyond vectorization nor algorithms for reducing complexity on large matrices, which limits the parameter values on which we improve performance.






\paragraph{System Setup}
MMM is implemented in C++ for shared memory parallel x86 CPUs.  We evaluate on a variety of x86 cpus from Intel (Table \ref{tab:cpus}).  We focus on Intel CPUs to compare with Intel's math kernel library, which has been highly optimized. The code was hand vectorized using Intel AVX intrinsics.  Two of the machines only support AVX2 while the others support AVX512.  The code is parallelized with \texttt{Cilk}~\cite{IntelCilkPlus10} and the Tapir~\cite{SchardlMoLe19, SchardlLe23} branch of the LLVM~\cite{Lattner02, LattnerAd04}
compiler (version 16).

\begin{table}[t]
  \centering
  \setlength{\tabcolsep}{4pt}
  \begin{tabular}{@{}lllrr@{}}\toprule
    \textit{Machine} & \textit{CPU}               & avx    & \textit{cpus} & \textit{threads} \\
    \midrule
    i9               & Intel Core i9-13900H       & AVX2   & 1             & 20               \\
    c4               & Intel Xeon E5-2666 v3      & AVX2   & 2             & 36               \\
    c5               & Intel Xeon Platinum 8275CL & AVX512 & 2             & 96               \\
    c6               & Intel Xeon Platinum 8375C  & AVX512 & 2             & 128              \\
    \bottomrule
  \end{tabular}%
  \caption{Computing platforms.}
  \label{tab:cpus}
\end{table}


\paragraph{Random Matrices}

Randomly generated matrices vary the random sparsity in $A$, fraction of zeros, and block random sparsity in $B$, fraction of 0 valued vectors.
Varying these quantities allows us to build maps that demonstrate speedup regions for MMM.

Figure \ref{fig:time-overview} shows the relative speedup to the Intel MKL on a consumer-grade machine (i9-13900H).
This experiment multiplies $2048^2$ matrices using all threads. We report the median of 10 trials.
We then run these same multiplications using the valgrind simulator and find that MMM executes significantly fewer instructions than either MKL approach in this range (Figure \ref{fig:inst-overview}). The reduction in instructions (4 times) exceeds the reduction in runtime (2 times), indicating that dense MKL realizes more instructions per cycle.

\begin{figure}
  \centering
  \begin{tikzpicture}
    \begin{axis}[
        enlargelimits=false, axis on top, axis equal image,
        xmin=0,
        xmax=1,
        ymin=0,
        ymax=1,
        xtick style={draw=none},
        ytick style={draw=none},
        xlabel = B Sparsity,
        ylabel = A Sparsity,
        ylabel near ticks,
        xlabel near ticks,
        ytick={0, .2, .4, .6, .8,1},
        yticklabels={1,.8,.6,.4,.2,0},
        axis x line*=top
      ]

      \addplot[thick,blue] graphics[xmin=0,ymin=0,xmax=1,ymax=1] {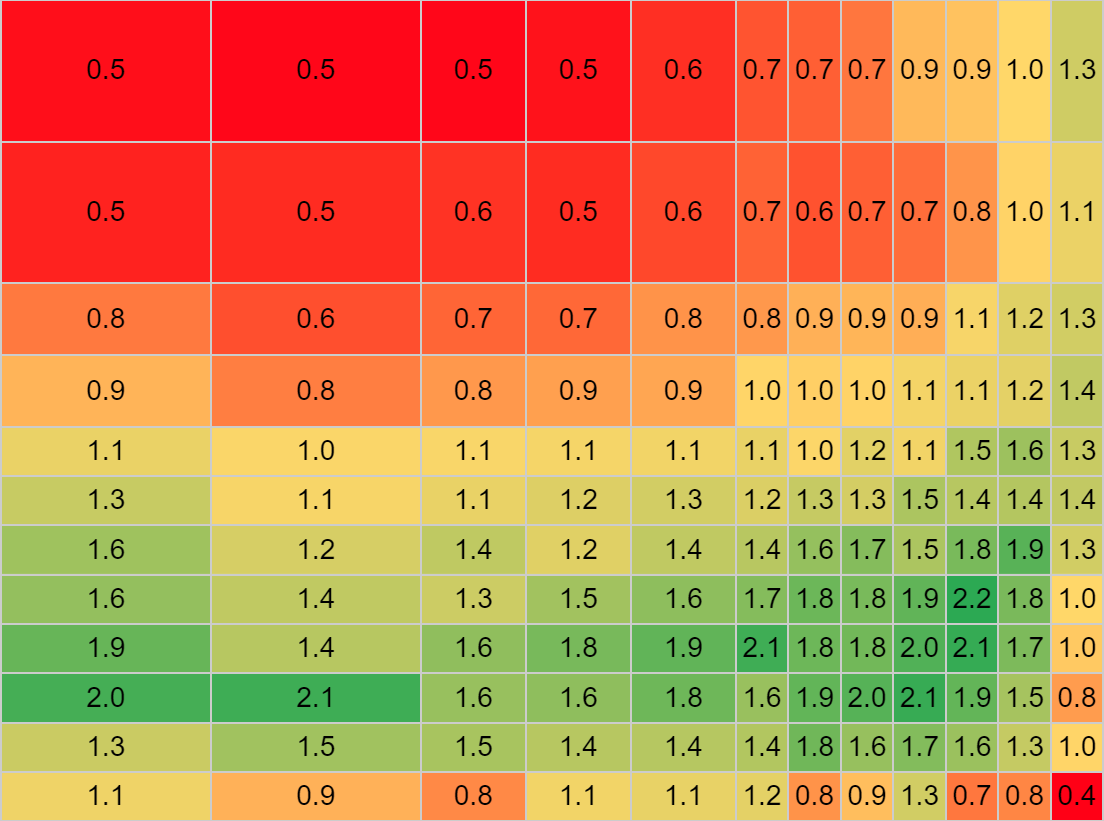};
    \end{axis}
  \end{tikzpicture}
  \caption{Heatmap of the performance of MMM versus Intel MKL's best algorithm. Numbers are speedup and green indicates better performance.}
  \label{fig:heatmap}
\end{figure}

\begin{figure}
  \centering
  \begin{tikzpicture}
    \begin{axis}[
        enlargelimits=false, axis on top, axis equal image,
        xmin=0,
        xmax=1,
        ymin=0,
        ymax=1,
        xtick style={draw=none},
        ytick style={draw=none},
        xlabel = B Sparsity,
        ylabel = A Sparsity,
        ylabel near ticks,
        xlabel near ticks,
        ytick={0, .2, .4, .6, .8,1},
        yticklabels={1,.8,.6,.4,.2,0},
        axis x line*=top
      ]

      \addplot[thick,blue] graphics[xmin=0,ymin=0,xmax=1,ymax=1] {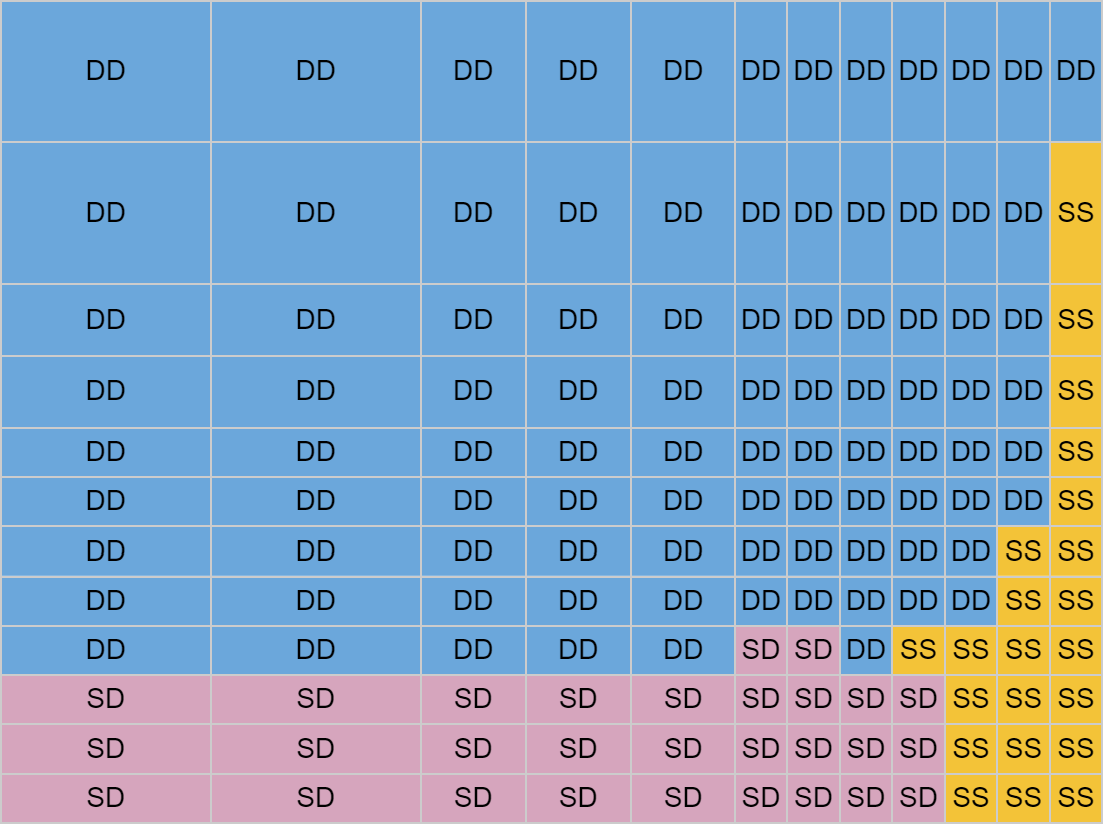};
    \end{axis}
  \end{tikzpicture}
  \caption{Intel's fastest multiplication algorithm--dense-dense (DD), sparse-dense(SD), and sparse-sparse(SS)--as a function of sparsity.}
  \label{fig:best-mkl}
\end{figure}

Figure \ref{fig:heatmap} shows the same speedup experiment over all values of $A$ and $B$ sparsity.
MMM provides speedup in configurations with $A$ sparsity higher than 0.5 and less than 0.95. The interesting region is when $B$ is also between 0.5 and 0.95 where we avoid computations in both matrices and
achieve a speedup of two times in many configurations. The region of high A sparsity and low B sparsity is less significant as all the benefit comes from skipping rows. Looking at the multiplication method chosen by Intel MKL reveals that MMM fits in an intermediate space between Intel MKLs different methods (Figure \ref{fig:best-mkl}). Our best speedups lie in the sparsity region in which MKL's optimizer switches from dense-dense, to sparse-dense, to sparse-sparse.


Our next experiment varies the size of the matrix to characterize preprocessing overheads and scaling.
Table \ref{tab:scaling_n} shows the performance on different sized matrices at 80\% sparsity in both $A$ and $B$.
Once the matrices are at least 2048 in size we achieve substantial speedup.
For small matrices, the $O(n^2)$ preprocessing to build function pointers consumes a larger fraction of overall $O(n^3)$ computing time and reduces MMM's speedup.
We note that the MMM approach only impacts the base multiplication of a single block and does not affect the high level algorithm or the parallelization strategy.

\begin{table}[t]
  \centering
  \setlength{\tabcolsep}{4pt}
  \begin{tabular}{@{}rrrrrrr@{}}\toprule
    \textit{n} & \textit{MMM} & \textit{mkl dense} & \begin{tabular}{@{}c@{}}\textit{\underline{mkl dense}} \\ \textit{MMM}\end{tabular} &  & \textit{mkl sparse} & \begin{tabular}{@{}c@{}}\textit{\underline{mkl sparse}} \\ \textit{MMM}\end{tabular} \\
    \midrule
    256        & 737          & 116                & 0.16                                                                                 &  & 207                 & 0.28                                                                                 \\
    512        & 1930         & 766                & 0.40                                                                                 &  & 688                 & 0.36                                                                                 \\
    1024       & 5922         & 6188               & 1.04                                                                                 &  & 8311                & 1.40                                                                                 \\
    2048       & 26613        & 52787              & 1.98                                                                                 &  & 47784               & 1.80                                                                                 \\
    4096       & 214471       & 327315             & 1.53                                                                                 &  & 320535              & 1.49                                                                                 \\
    8192       & 1608555      & 2511919            & 1.56                                                                                 &  & 2877960             & 1.79                                                                                 \\
    \bottomrule
  \end{tabular}%
  \caption{Runtime (s) and speedup that MMM achieves with different matrix sizes for a sparsity of .8. Larger matrices amortize preprocessing costs in MMM.}
  \label{tab:scaling_n}
\end{table}

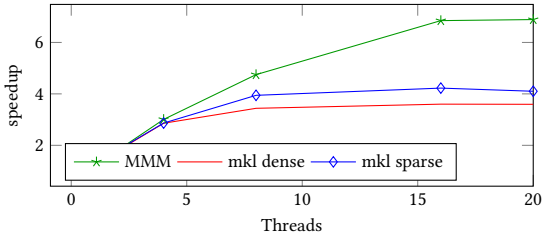
\begin{figure}
  \centering
  \footnotesize
  \begin{tikzpicture}
    \begin{axis}[height=4cm, xmax=20,
        width=8cm,        legend pos=south west,
        legend columns=3,        xlabel = Threads,
        ylabel = speedup,ylabel near ticks, xlabel near ticks]
      \addplot[mark size=2pt, mark=star, mygreen] table [x=alg, y=MMM, col sep=tab] {figures/8192_scaling.tsv};
      \addlegendentry{MMM}
      \addplot[mark size=2pt, mark=circle, red] table [x=alg, y=mkl dense, col sep=tab] {figures/8192_scaling.tsv};
      \addlegendentry{mkl dense}
      \addplot[mark size=2pt, mark=diamond, blue] table [x=alg, y=mkl sparse, col sep=tab] {figures/8192_scaling.tsv};
      \addlegendentry{mkl sparse}

    \end{axis}
  \end{tikzpicture}
  \caption{Multicore scalability of MMM on $8192^2$ matrices.}
  \label{fig:scaling_p}
\end{figure}

We present the scalability of MMM from 1 to 20 threads (Figure \ref{fig:scaling_p}).
MMM scales better. However, MKL starts slower at 1 thread. MMM is faster than MKL only at 8 or more threads.
We hypothesize that MMM requires concurrency to overlap memory accesses and computation. That is, that
sparse regions lead to memory-bound, low-compute periods in each thread.


\paragraph{Server Architectures}

We run experiments on three generations of Intel server processors to show the potential of MMM on different architectural and memory configurations. These experiments both (1) show that MMM reduces instruction counts and (2) verify that there are parameters and configurations on which MMM provides speedup. There are many limitations to using whole, multi-socket machines with MMM's current implementation. Most importantly, MMM doesn't have optimizations for large matrices.  To create a workload that uses the entire machine and keeps matrices mid-size, we run $n$ independent processes, each of which runs a few matrix multiplies and sum up the total time.  This roughly corresponds to a transformer/MLP workload in which the dimensionality of any operation is limited by the layer. MMM does not work well on single massive matrix multiplies that would use the entire machine.

Table \ref{tab:run_many} shows that MMM achieves reasonable speedups on all three generations of CPUs and does better when computing floats than doubles.  We also measure the number of instructions each computation takes and report the ratios.  MMM performs many fewer instructions that MKL dense in all cases. We even perform fewer instructions than MKL sparse in all cases. This is just the multiplication and does not include conversion costs,~i.e.~MKL starts with sparse matrices. Instruction counts were measured using performance counters on the c5 and c6 and valgrind on the C4 (did not finish in larger cases). Additionally, we find that the instruction count reduction is substantially higher on the c4 in AVX2 with 256-bit vectors. MMM does not yet fully exploit the additional registers available to AVX-512.


\begin{table}[t]
  \centering
  \setlength{\tabcolsep}{4pt}
  \begin{tabular}{@{}lrlrrrr@{}}\toprule
    \textit{Machine} & \textit{Size} & \begin{tabular}{@{}c@{}}\textit{data} \\ \textit{type}\end{tabular} & \textit{Sparsity} & \textit{speedup} & \begin{tabular}{@{}c@{}}\textit{\underline{I dense}} \\ \textit{I MMM}\end{tabular} & \begin{tabular}{@{}c@{}}\textit{\underline{I sparse}} \\ \textit{I MMM}\end{tabular} \\
    \midrule
    c4               & 2048          & float                                                               & 0.88              & 1.26             & 6.75                                                                                   & 2.18                                                                                   \\
    c4               & 2048          & double                                                              & 0.86              & 1.33             & 7.69                                                                                   & 2.14                                                                                   \\
    c4               & 4096          & float                                                               & 0.88              & 1.27             & DNF                                                                                    & DNF                                                                                    \\
    c4               & 4096          & double                                                              & 0.85              & 1.25             & DNF                                                                                    & DNF                                                                                    \\
    \midrule
    c5               & 2048          & float                                                               & 0.9               & 1.12             & 5.02                                                                                   & 1.35                                                                                   \\
    c5               & 2048          & double                                                              & 0.88              & 1.29             & 6.06                                                                                   & 1.29                                                                                   \\
    c5               & 4096          & float                                                               & 0.92              & 1.21             & 7.47                                                                                   & 1.10                                                                                   \\
    c5               & 4096          & double                                                              & 0.88              & 1.35             & 7.67                                                                                   & 1.18                                                                                   \\
    \midrule
    c6               & 2048          & float                                                               & 0.91              & 0.80             & 4.66                                                                                   & 1.14                                                                                   \\
    c6               & 2048          & double                                                              & 0.89              & 1.11             & 6.68                                                                                   & 1.13                                                                                   \\
    c6               & 4096          & float                                                               & 0.93              & 0.82             & 8.13                                                                                   & 1.13                                                                                   \\
    c6               & 4096          & double                                                              & 0.88              & 1.22             & 7.83                                                                                   & 1.21                                                                                   \\
    \bottomrule
  \end{tabular}%
  \caption{Best speedup (varying sparsity) over MKL for each machine, size, data type combination. $\frac{\text{I dense}}{\text{I MMM}}$ and $\frac{\text{I sparse}}{\text{I MMM}}$ gives instruction count ratios for MKL dense MKL sparse respectively. This was measured using valgrind for the c4 and performance counters for c5 and c6.}
  \label{tab:run_many}
\end{table}

\paragraph{Matrices from ML workloads}

We measure MMM's performance on matrices gathered from transformer inference after sparsification of the MLP layers in Deja Vu \cite{liu2023deja}.  We used the 1.3 billion parameter Open Pre-Trained Transformer Language (OPT 1.3b) from Hugging Face~\cite{zhang2022opt}. The MLP layer computes:
\begin{equation}
  MLP_{SM}(Y) = \sigma(yW^{1}_{SM})(W^{2}_{SM})^T
\end{equation}
in which $\sigma$ is the GeLU activation function. $W^1, W^2 \in R^{d \times 4d}$ and $d=2048$. $SM \subseteq [4d]$ is the set of neurons that are activated. The remaining neurons are zero resulting in column sparse matrices. The first matrix $\sigma(yW^{1}_{SM})$ is column sparse.
The second matrix ($(W^{2}_{SM})^T$) is row sparse since it is the transpose of a column sparse matrix.

This is one specific instance of sparse-sparse multiplication in Deja Vu. There are many sparsified matrices that we do not use, including column-sparse transformer heads and random sparse activations. However, the block sparse attention heads are multiplied by a projection matrix that is dense. The projection matrix cannot be sparsified without losing information because it combines information from all the attention heads.  
We expect many more operations to have two-sided sparsity, but that requires integrating sparsity in the model design~\cite{wang2021dual}.

MMM achieves speedup on many of the matrices from Deja Vu's MLP layers. The degree of speedup depends on the sparsity parameter, which Deja Vu varies. Table \ref{tab:ml_mats} shows three examples at a favorable sparsity (88\%) within Deja Vu's target range.

\begin{table}[t]
  \centering
  \setlength{\tabcolsep}{2pt}
  \begin{tabular}{@{}crrrr@{}}\toprule
    \textit{A sparsity} & \textit{B sparsity} & MMM   & MKL SD & speedup \\
    \midrule
    0.88                & 0.88                & 13102 & 19678  & 1.5     \\
    0.88                & 0.88                & 14247 & 20104  & 1.4     \\
    0.88                & 0.88                & 13045 & 20065  & 1.5     \\

    \bottomrule
  \end{tabular}%
  \caption{Speedup over MKL sparse on three matrices from ML workloads. Sparse-dense was the fastest MKL approach.}
  \label{tab:ml_mats}
\end{table}


\section{Related Work}

Nvidia sparse tensor cores~\cite{mishra2021accelerating} pursue a similar goal of mapping sparse computations onto dense matrix hardware. They map 50\% sparse computations onto dense hardware using a 2:4 strategy, operating on two non-zero elements of 4 adjacent elements. This is effective on a wide-variety of models that can be sparsified so that a 2:4 encoding preserves the accuracy of the dense model. The 2:4 principle has been generalized to M:N sparsity and further work shows how to quantify performance loss, prune models, and use M:N sparsity during training~\cite{zhou2021learning, Sun2021DominoSearchFL}. MMM makes no assumptions about fine-grained sparsity.

The Tensor Algebra compiler (TACO)~\cite{taco} produces optimized computation kernels for general tensor operations on many sparse data formats. Techniques have been extended to GPUs~\cite{Senanyake20sparse}, distributed memory and heterogeneous architectures~\cite{yadav2022distal}, and dynamic data structures that support updates~\cite{chou2022dynamic}. Compiled kernels match or approach the performance of hand-optimized code, easing development effort and making code robust to changing data and computing environments. Tensor compilation could be extended to use just-in-time compilation or dynamic code evaluation to adapt to sparsity at runtime. However, all compilation techniques target a sparse data structure. MMM targets mid-ranges of sparsity in which there are not enough zeros to justify sparse data structures as demonstrated by our performance against Intel's sparse-sparse and sparse-dense methods. Also, compiled kernels are arbitrarily complex and MMM supports only GeMM.

The principles of tensor compilation and sparse data structures have been extended to the design of sparse accelerators. Dual-Side Sparse Tensor Core~\cite{wang2021dual} uses bitmap representations to perform outer-product SpGEMM in hardware. It captures sparsity in both activations and weights. SparseLoop~\cite{wu2021sparseloop} builds a cost-driven optimizer that generalizes the principles in DSTC and NVidia 2:4. It generates designs for multiple sparse representations and compression formats. An abstract dataflow machine model for sparse tensor algebra leads to a compiler and design of reconfigurable and fixed-function accelerators~\cite{hsu2023abstract}.



\section{Conclusion and Discussion}
\balance

MMM shows the potential to exploit emergent sparsity for matrix multiplication on transformers.  It demonstrates a reduction of the work needed to perform matrix multiplication on sparse matrices.

Future work involves involves improvements to more classes of matrices such as larger or denser, as well as extend it to work on different architectures.
The low-level code for multiplying small regions could be improved to increase the register reuse and could outperform MKL dense for a larger range of sparsities.  Other algorithmic style optimizations could be used such as using a JIT to compile the exact forms needed for the existing patterns, instead of pre-compiling all possible forms.
The code could also be extended to seamlessly choose the optimal base function for each recursive block between both MMM approaches as well as different standard approaches from a vendor library.

We plan to extend this work to GPUs.  This will involve some changes to the required sparsity structure.  For example, to maximize the performance and align the different warps, all of the threads in a warp need to compute on the same pattern of rows of B and elements from A.  This would require vertical blocks in A in addition to horizontal blocks in B. The $A$ matrices of MLP computation were column matrices and implement this pattern. We believe this approach will extend to GPUs.








\bibliographystyle{ACM-Reference-Format}
\bibliography{sample-base}


\end{document}